\documentclass[twocolumn, a4paper,usenatbib]{aastex6}
\usepackage[T1]{fontenc}
\usepackage{ae,aecompl,graphicx,amsmath,amssymb,bm,subfig,array,booktabs,times} 

\newcommand{\ie}{i.e.,~}
\newcommand{\eg}{e.g.,~}

\begin{document}

\title{A General-relativistic Determination of the Threshold Mass to
   Prompt Collapse in Binary Neutron Star Mergers}

\newcommand{\ouraffiliation}{%
	\affiliation{Institut f{\"u}r Theoretische Physik, Goethe-Universit{\"a}t
        Max-von-Laue-Stra{\ss}e 1, 60438 Frankfurt, Germany}
}

\author{Sven~K{\"o}ppel}\ouraffiliation
\author{Luke~Bovard}\ouraffiliation
\author{Luciano~Rezzolla}\ouraffiliation

\begin{abstract}
We study the lifetimes of the remnant produced by the merger of two
neutron stars and revisit the determination of the threshold mass to
prompt collapse, $M_{\rm th}$. Using a fully general-relativistic
numerical approach and a novel method for a rigorous determination of
$M_{\rm th}$, we show that a \emph{nonlinear} universal relation exists
between the threshold mass and the maximum compactness. For the
temperature-dependent equations of state considered here, our results
improve a similar \emph{linear} relation found recently with methods that
are less accurate but yield quantitatively similar results. Furthermore,
exploiting the information from GW170817, we use the universal relation
to set lower limits on the stellar radii for any mass.
\end{abstract}

\maketitle

\section{Introduction}
\label{sec:intro}

The recent detection of gravitational waves from the merger of
neutron-star binaries~\citep{Abbott2017_etal} has heralded the new era of
multi-messenger gravitational-wave astronomy. These observations offer
new insight into the most extreme objects in universe, namely, neutron
stars, and allow us to probe and constrain the properties of nuclear
matter \citep{Annala2017, Paschalidis2017, Bauswein2017b, Radice2017b,
  Most2018, Burgio2018, Montana2018}.

When two neutron stars merge, they will produce an object that either
collapses promptly to a black hole, or does not \citep{Baiotti08}. In the
latter case, the remnant may be a metastable object, \eg a hypermassive
neutron star (HMNS), eventually collapsing to a black hole on a secular
timescale, or survive for much longer times, either as a rotating or a
nonrotating star \citep[see, \eg][for a review]{Baiotti2016}. In the case
of the first detection of merging neutron stars, GW170817
\citep{Abbott2017_etal}, the precise fate of the merger remnant is
presently unknown, although the formation of a black hole naturally
matches the simultaneous observation of a short gamma-ray burst
\citep{Eichler89, Rezzolla:2011}, and has been the working hypothesis to
set new limits on the maximum mass of neutron stars \citep{Margalit2017,
  Shibata2017c, Rezzolla2017, Ruiz2017}.

Determining the time of collapse of the merger remnant is particularly
challenging as there are a number of physical processes that either
determine or undermine the stability of merger remnant. These include:
the ejection of matter \citep{Rosswog1999, Kyutoku2012, Radice2016,
  Lehner2016, Dietrich2016, Bovard2017}, the angular-momentum transfer
via magnetic fields \citep{Kiuchi2015a, Siegel2013, Kawamura2016}, the
evolution of the degree of differential rotation \citep{Kastaun2016,
  Hanauske2016}, and possible viscous effects mediated either by
neutrinos or magnetic fields \citep{Duez2004b, Shibata:2017b, Radice2017,
  Alford2017}.

The determination of the critical (threshold) mass to a prompt collapse,
$M_{\rm th}$, is much simpler, although it still poses numerical and
conceptual challenges. \citet{Bauswein2013} have been the first to
explore this problem by employing a smooth-particle approximation for the
hydrodynamics and a conformally flat approximation to general
relativity. In this way, they were able to find a \emph{linear} universal
relationship between $M_{\rm th}$ and the compactness of the maximum-mass
model, $\mathcal{C}_{_{\rm TOV}}:=M_{_{\rm TOV}}/R_{_{\rm TOV}}$, where
$M_{_{\rm TOV}}$ and $R_{_{\rm TOV}}$ are respectively the mass and
radius of the maximum-mass nonrotating star. Here, we improve on this
result by using a fully general-relativistic approach, a wider range of
compactnesses, and a rigorous definition of the threshold mass. As a
result, we find a \emph{nonlinear} relation between $M_{\rm th}$ and
$\mathcal{C}_{_{\rm TOV}}$, which offers a better match to the
numerical-relativity results. Furthermore, exploiting the information
from GW170817, we use the new relation to set more stringent lower bounds
on the radii neutron stars.

\newpage
\section{Methods and setup}
\label{sec:methods}

To describe the evolution of the merging system, we solve the coupled
Einstein-hydrodynamics system~\citep{Rezzolla_book:2013} using the
\texttt{Ein\-stein\- Toolkit}~\citep{loeffler_2011_et}. In particular, we
evolve the spacetime with the \texttt{McLachlan}
code~\citep{Brown:2008sb}, with the same gauges as
in~\citet{Hanauske2016}. On the other hand, we evolve the matter with the
high-order relativistic-hydrodynamics code
\texttt{WhiskyTHC}~\citep{Radice2012a, Radice2013c}. The numerical grid
uses the fixed-mesh refinement driver \texttt{Carpet}
\citep{Schnetter-etal-03b}, with a total of six refinement levels having
a highest resolution of $215\,{\rm m}$ covering the two stars and a total
extent of $700\,{\rm km}$. For one equation of state (EOS), we also
considered different resolutions of $215, 287$ and $573\,{\rm m}$,
obtaining threshold masses within a variance of $\Delta M_{\rm th}
\lesssim 0.005\,M_{\odot}$.

\begin{figure}
	\includegraphics[width=\linewidth]{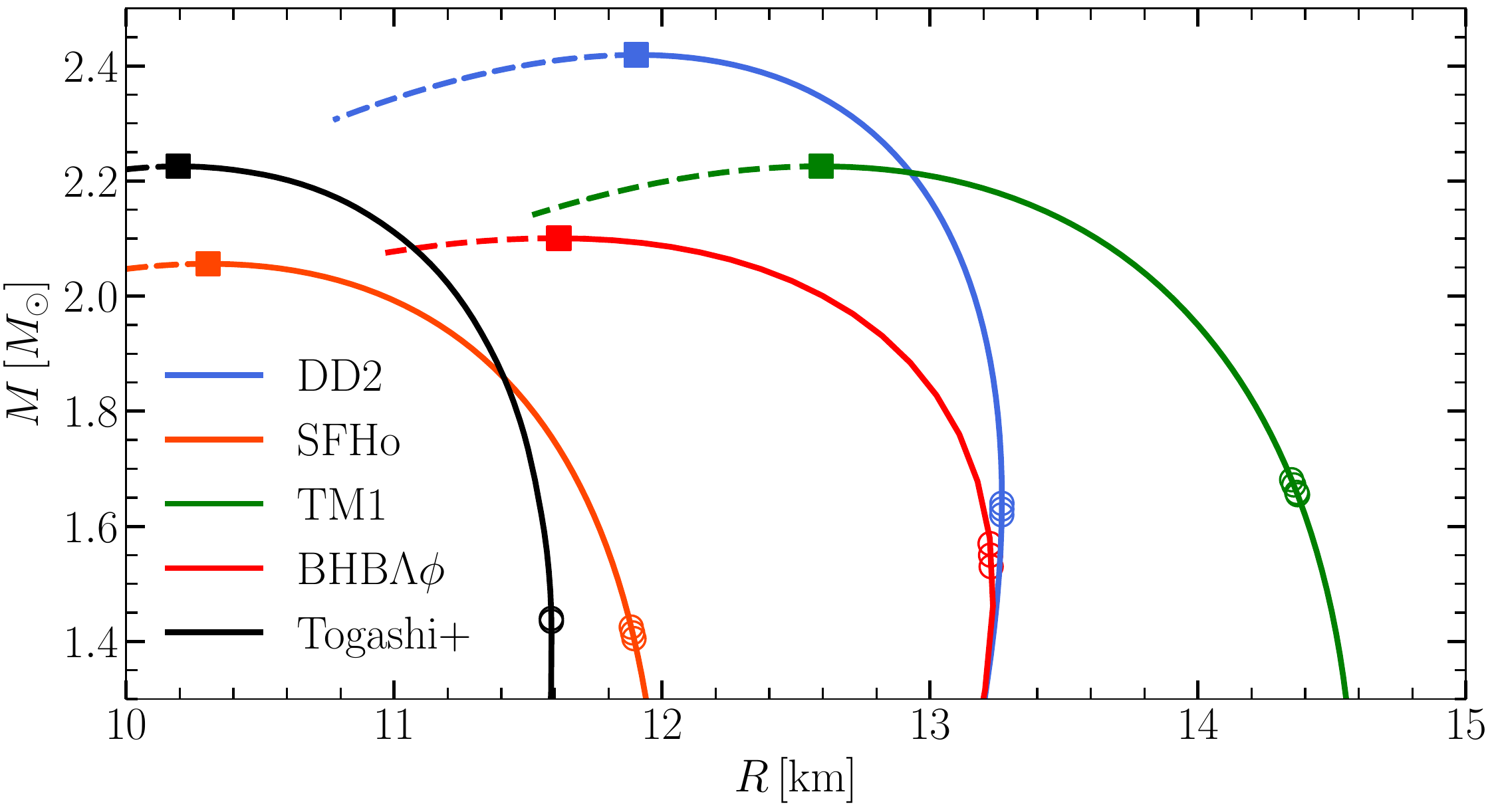}
	\caption{Masses and radii of nonrotating equilibrium solutions,
          both stable (solid lines) and unstable (dashed lines). Solid
          squares mark the maximum-mass solutions, while open circles
          refer to models used as initial data; note that some EOSs have
          initial data with similar properties but different maximum
          masses.}
	\label{fig:eos-properties}
\end{figure}%

A number of zero-temperature (``cold'') EOSs are available for numerical
simulations and rely on nuclear-physics calculations following a variety
of theoretical approaches. While these EOSs are suitable to describe the
inspiral phase, they become obviously inadequate after the merger, when
the temperatures reach values of several tens of MeV. To counter this, it
is customary to model the post-merger dynamics by modifying these EOS and
adding a ``thermal'' contribution via an ideal-fluid EOS
\citep{Rezzolla_book:2013} so as to account for the shock heating
\citep{Janka93}. This approach, while not self-consistent, is rather
robust and such ``hybrid EOSs'' have been employed extensively in the
literature \citep{Baiotti2016}.

\begin{table}
  \centering
  \begin{tabular}{lcccccc}
    \toprule
    EOS & $M_{_{\rm TOV}}$ & $R_{_{\rm TOV}}$ & $C{_{\rm TOV}}$ & 
    $\tau_{_{\rm TOV}}$  & $M_{\rm th}$ & $\Delta M_{\rm th}$ \\
    & \footnotesize $[M_{\odot}]$ &  \footnotesize $[{\rm km}]$ &  \footnotesize  
    & \footnotesize  \footnotesize   $[\mu 
      {\rm s}]$ &  \footnotesize  $[M_{_{\rm TOV}}]$ & \footnotesize  $[M_{_{\rm TOV}}]$  \\
    \midrule 
    BHB$\Lambda\Phi$  &           2.10 &        11.64 &   0.26 & 83.31 & 
    1.503 & 0.005 \\
    DD2               &           2.42 &        11.94 &   0.30 & 80.60 & 
    1.364 & 0.020  \\
    SFHo              &           2.06 &        10.34 &   0.29 & 70.44 & 
    1.391& 0.016 \\
    TM1               &           2.22 &         12.6 &   0.26 & 91.70 & 
    1.520 & 0.015 \\
    Togashi$+$        &           2.23 &        10.17 &   0.32 & 66.12 & 
    1.298 & 0.000\\
    \botrule
  \end{tabular}
  \caption{Properties of the maximum-mass models for the EOSs considered
    here: BHB$\Lambda\Phi$~\citep{Banik2014},      
    \mbox{(HS-)}DD2~\citep{Typel2010},
    SFHo~\citep{Steiner2013}, (HS-)TM1~\citep{Hempel2012}, and
    Togashi$+$~\citep{Togashi2016}. Reported are: the maximum mass of a
    nonrotating star $M_{_{\rm TOV}}$, the corresponding radius $R_{_{\rm
        TOV}}$, the compactness $C_{_{\rm TOV}}$, and the
    free-fall-timescale $\tau_{_{\rm TOV}}$. Also shown are the threshold
    masses $M_{\rm th}$ and the corresponding errors.}
  \label{table:eos-properties}
\end{table}

However, since our goal here is that of determining as accurately as
possible the threshold mass to prompt gravitational collapse, it is
essential that the description of the thermal effects in the matter is as
realistic and self-consistent as possible. In turn, this forces us to
consider EOSs that have a physically consistent dependence on
temperature. Unfortunately, the number of EOSs that can be employed for
this scope and that do not violate some basic nuclear-physics requirement
\citep[as it is the case for the widely employed LS220
  EOS,][]{Kolomeitsev2016}, is much more restricted. Here, we have
employed all of the five ``hot'' EOSs that can be used with confidence to
determine the threshold mass. The corresponding properties, when
expressed in terms of the masses and radii of the maximum-mass
nonrotating configuration (hereafter TOV), are reported in
Table~\ref{table:eos-properties}. Similarly,
Fig.~\ref{fig:eos-properties}, shows the masses and radii of the TOV
equilibrium solutions, both stable (solid lines) and unstable (dashed
lines). Solid squares mark the maximum-mass solutions, while open circles
the models used.

Using these EOSs, we have modelled the initial data under the assumption
of irrotational quasi-circular equilibrium
\citep{Gourgoulhon-etal-2000:2ns-initial-data} and computed it via the
\texttt{LORENE} library for a total of 15 equal-mass (\ie $q=1$)
binaries. The initial separation is of $45 \,{\rm km}$, so that the
binaries perform around five orbits before the merger. We note that since
the threshold mass for equal-mass binaries is always larger than for
unequal-mass binaries, \ie $M_{\rm th}(q=1)>M_{\rm th}(q<1)$, the use of
equal-mass binaries is not a restriction but optimises the search for
$M_{\rm th}$~\citep[see also][for a discussion]{Bauswein2017b}.

The measurement of the threshold mass inevitably imposes a clear
definition, but also a straightforward procedure to extract this
information from the numerical-relativity simulations. Quite generically,
one expects that the lifetime of the HMNS, or collapse time $t_{\rm
  coll}$, will decrease as the mass of the binary is increased, so that
the threshold mass will represent the shortest possible lifetime. This
definition, is however inconvenient since different EOSs will yield
different ``shortest lifetimes'' and comparing different EOSs may
introduce clear biases. We resolve this problem by building our analysis
around two important logical steps. First, we consider the collapse time
as a dimensionless quantity by expressing it in terms of the free-fall
time $\tau_{\rm ff}$ that, for an object of mass $M$ and radius $R$, is
given by \citep{Rezzolla_book:2013}
\begin{equation}
  \label{eq:ff}
  \tau_{\rm ff} (M,R) := \frac \pi 2 \sqrt{\frac{R^3}{2M} } \,.
\end{equation}
Since $R\sim 1/M$ for stable models, the smallest free-fall time will be
achieved for the maximum-mass model, so that the shortest free-fall
timescale is $\tau_{_{\rm TOV}}:=\tau_{\rm ff}(M_{_{\rm TOV}},R_{_{\rm
    TOV}})$. Second, we define the threshold mass $M_{\rm th}$ as the one
for which the merger remnant will collapse over such a timescale, \ie
$M/M_{_{\rm TOV}}\to M_{\rm th}/M_{_{\rm TOV}}$ for $t_{\rm
  coll}/\tau_{_{\rm TOV}}\to 1$.

We next discuss the procedure followed for the measurement of the
collapse time, $t_{\rm coll}$ as given by the difference between the
(coordinate) time of \textit{merger}, $t_{\rm merg}$, and that of
\textit{collapse} to a black hole, $t_{_{\rm BH}}$, \ie $t_{\rm
  coll}:=t_{_{\rm BH}}-t_{\rm merg}$. Both of these times can be measured
in a number of different ways. The first one involves the emission of
gravitational waves, with $t_{\rm merg}$ being given by the time of the
first maximum of the gravitational-wave strain amplitude $h$, or of the
Weyl scalar $\Psi_4$. The time of collapse, on the other hand, can be
estimated as the time when the ringdown signal starts. While overall
robust, measurements involving gravitational waves are prone to errors as
$t_{\rm merg}$ depends sensitively on the phase evolution of the binary
in its most nonlinear stage\footnote{As the threshold mass is approached,
  the merger can take place a fraction of a radian earlier/later than
  expected, biasing the measurement.}. Similarly, determining $t_{_{\rm
    BH}}$ is complicated by the fact that the beginning of the ringdown
is somewhat arbitrary and ringdown itself can be modified by the
matter infalling onto the black hole.

To counter these difficulties, $t_{\rm merg}$ could be measured via the
proper separation between the two stellar cores, marking the merger as
the time when such a separation is below a fraction of the initial
diameters of the two stars. Similarly, $t_{_{\rm BH}}$ could also be
measured in terms of the first appearance of an apparent horizon.
Although these measurements show an overall consistency, they suffer in
from the fact that the proper separation is very sensitive to the
properties of the EOSs, and that the first appearance of the apparent
horizon is ultimately set by the frequency at which it is searched during
the simulations.

A third approach for measuring $t_{\rm coll}$ and $t_{_{\rm BH}}$
involves instead the monitoring of the minimum of the lapse function
$\alpha$, which we evolve employing a singularity-avoiding ``$1+\log$''
slicing~\citep{Alcubierre:2008}. This quantity has been shown to be a
very good proxy for the tracking and appearance of an apparent horizon
\citep{Alcubierre:2008} and has the advantage of being extremely
robust. As a result, we can mark the two times respectively when
\begin{equation}
  \label{eq:arbitrary-alpha-values}
  \begin{aligned}
    & t_{\rm merg}:& &\min(\alpha) = \alpha_\textrm{merg} := 0.35\,,\\
    & t_{_{\rm BH}}:& &\min(\alpha) = \alpha_{_{\rm BH}} := 0.2 \,.
  \end{aligned} 
\end{equation}
These values for $\min(\alpha)$ are less arbitrary than they may appear
as $\alpha_{\rm merg}=0.35$ and $\alpha_{_{\rm BH}}=0.2$ systematically
represent the first minimum and zero of second derivative of the function
$\min(\alpha(t))$, respectively; also, at $\alpha_{_{\rm BH}}=0.2$ the
apparent horizon is also normally first found. More importantly, the
results are invariant under a change of the values in
\eqref{eq:arbitrary-alpha-values}.

In summary, all methods to compute $t_{\rm coll}$ provide results that
are consistent across different choices. However, when it comes to
robustness and simplicity of implementation, the monitoring of the
minimum of the lapse function represents the optimal choice and is the
one employed in the results that will be presented next.

\section{Results}
\label{sec:results}

\begin{figure}
  \includegraphics[width=\linewidth]{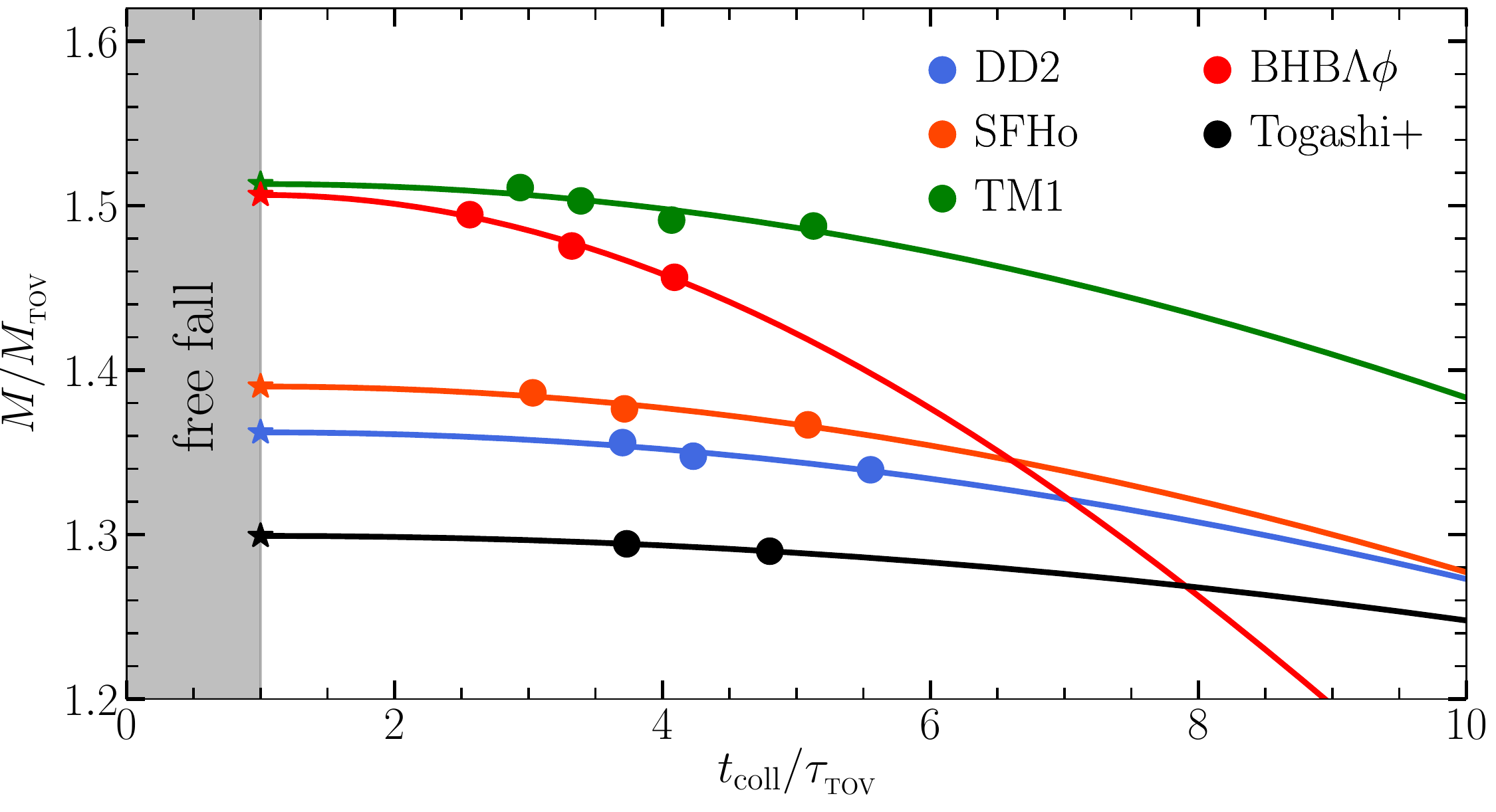}
  \caption{Measured collapse times $t_{\rm{coll}}$ (circles) normalised
    by their respective $\tau_{_{\rm{TOV}}}$ for the different
    EOSs. Stars represent the threshold mass, $M_{\rm{th}}$, predicted by
    the exponential fit, while the grey-shaded region corresponds to
    times below free-fall.}
\label{fig:lin_noscale}
\end{figure}

Figure~\ref{fig:lin_noscale} reports the collapse times normalised to the
free-fall timescale of the maximum-mass models, $t_{\rm coll}/\tau_{_{\rm
    TOV}}$, for all of the EOSs considered; these times are then shown as
a function of the initial half-mass of the binary normalized to the
maximum-mass, $M/M{_{\rm TOV}}$. The adoption of such set of
dimensionless quantities has the goal of revealing a behaviour of the
threshold mass that is universal, \ie only weakly dependent on the EOS
~\citep[see, \eg][for some examples]{Breu2016, Weih2017}.

Filled circles of different colours in Fig.~\ref{fig:lin_noscale} report
the numerical data for the various EOSs and mark the values of the
initial masses in the binaries. Note that as $t_{\rm{coll}}$ decreases,
even small differences in the initial masses can lead to rather large
differences in the survival time. Also note that only two values are
reported for the Togashi$+$ EOS and these differ by only $3.7\%$ in mass
(\ie $M=1.440,\,1.435\,M_{\odot}$); a binary with a slightly smaller mass
(\eg $M=1.430\,M_{\odot}$) leads to a HMNS effectively stable over the
timescales investigated here. Finally, since $t_{\rm{coll}}$ should
diverge for vanishingly small values of $M$, we fit the numerical data
with a simple exponentially decaying function of the type
$M/M_{_{\rm{TOV}}} = \tilde{a}\exp[-\tilde{b} ( t_{\rm coll}/\tau_{_{\rm
      TOV}} )^{2}]$. A rapid inspection of Fig.~\ref{fig:lin_noscale}
shows that the exponential fit is rather good and that a linear
approximation would overestimate the threshold mass. Indeed, near the
free-fall limit, the behaviour of $M/M_{_{\rm{TOV}}}$ should not be
linear as in this limit infinitesimal changes in $M$ are sufficient to
yield a prompt collapse. Such a behaviour, frequently encountered in
critical-collapse calculations, requires the function $M/M_{_{\rm{TOV}}}$
to have vanishing derivative for $t_{\rm coll}/\tau_{_{\rm TOV}} \to
1$. Clearly, our nonlinear fitting reflects this behaviour while a linear
one does not.

Figure~\ref{fig:lin_noscale} also reveals that the threshold mass is
roughly given by
\begin{equation}
\label{eq:rough_fit}
  \frac{M_{\rm th}}{M_{_{\rm TOV}}} \approx 1.415\,,
\end{equation}
with an uncertainty of $\Delta M_{\rm th}=0.05 M_{\odot}$, \ie a relative
error of $\sim4\%$. Hence, Eq.~\eqref{eq:rough_fit} provides the
lowest-order approximation between the threshold mass and the
corresponding maximum mass. The existence of a relation of this type has
been suggested initially by~\citet{Bauswein2013}, who concluded that
${M_{\rm th}}/{M_{_{\rm TOV}}}=k$, with $k$ a \emph{linear} function of
the maximum compactness $\mathcal{C}_{_{\rm TOV}}$. More specifically,
the universal ansatz proposed by \citet{Bauswein2013} is that
$k=\hat{a}\,\mathcal{C}_{_{\rm TOV}}+\hat{b}$, with
$\hat{a}=3.38,\,\hat{b}=2.43$, independent of the EOS.

Such a linear ansatz does represent a reasonable first approximation to
the data but not the most general one. In particular, if it represents a
universal behaviour of compact self-gravitating objects it should be
valid for all possible compactnesses and provide the expected black-hole
limit, for which ${M_{\rm th}}/{M_{_{\rm TOV}}}\to0$ for
$\mathcal{C}_{_{\rm TOV}}\to1/2$. Hence, we correct the linear
approximation via a \emph{nonlinear} fit of the type
\begin{align}
  \label{eq:fit}
  \frac{M_{\rm th}}{M_{_{\rm TOV}}} = a - \frac{b}{1 -
    c\,\mathcal{C}_{_{\rm{TOV}}}}\,,
\end{align} 
where $a,b,c$ are to be determined from the data. However, imposing the
fulfilment of the black-hole limit removes one free parameter and sets
$a=2b/(2-c)$.

\begin{figure}
  \includegraphics[width=\columnwidth]{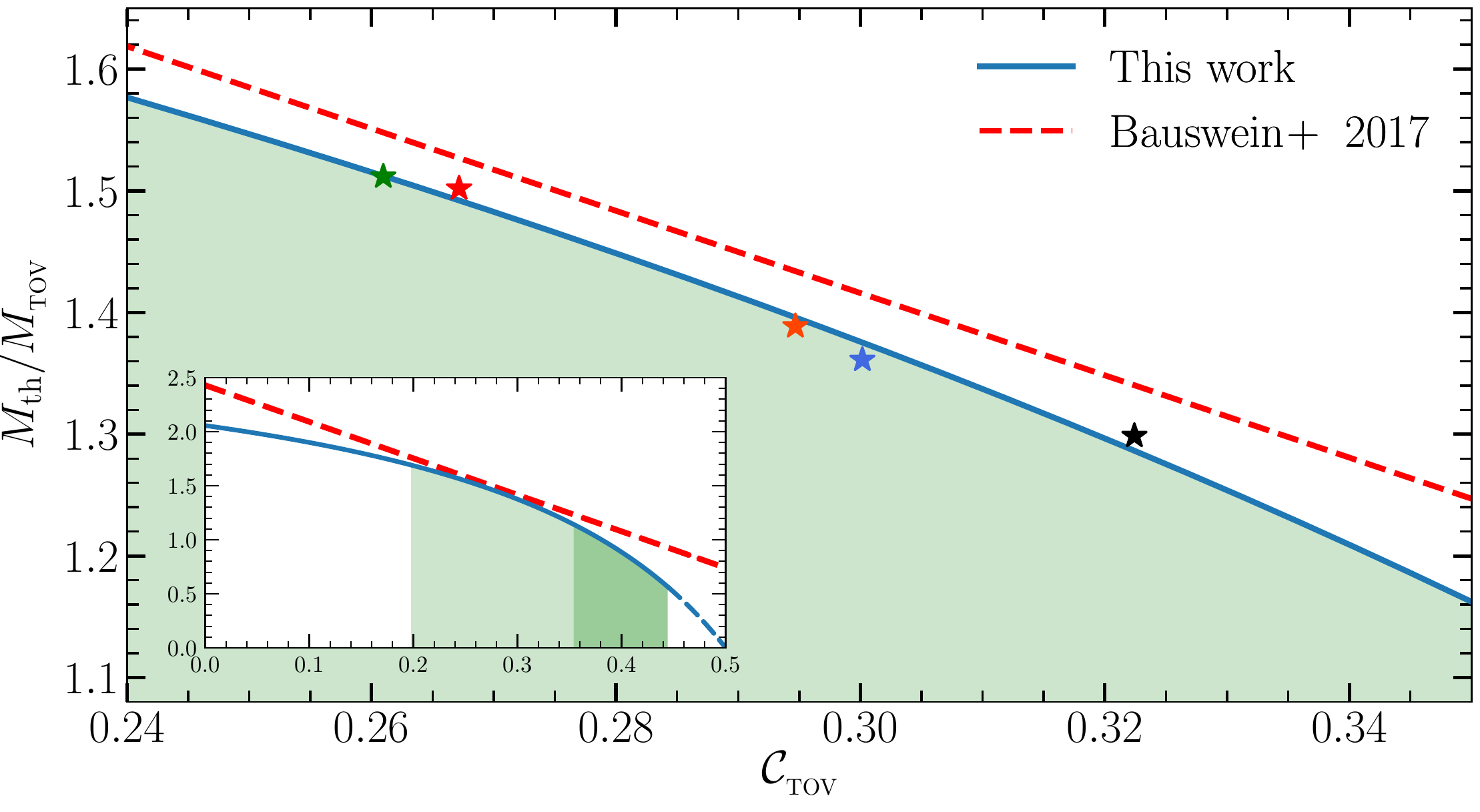}
    \caption{Universal relation for the threshold mass with the stars
      matching the data in Fig.~\ref{fig:lin_noscale}; the solid-blue
      line is the nonlinear fit (\ref{eq:fit}), while red-dashed line the
      linear fit of~\citet{Bauswein2017b}. The green-shaded area reports
      the compactness expected for neutron stars, while the inset the two
      universal relations over a wider range.}
    \label{fig:uni_rel}
\end{figure}

Figure~\ref{fig:uni_rel} reports with a solid blue line the fit
\eqref{eq:fit} with $b = 1.01,\,c =1.34$, against the
numerical-relativity data shown with stars of the same colorcode as in
Fig.~\ref{fig:lin_noscale} (see Table~\ref{table:eos-properties} for the
errors in the fit). Also shown with a red-dashed line is the linear
approximation of~\citet{Bauswein2017b}, which clearly suggests larger
threshold masses, most probably because the conformally flat
approximation used by \citet{Bauswein2013} underestimates the
strong-curvature behaviour that characterises the threshold to black hole
collapse. At the same time, the difference with the linear approximation
of is not enormous and is of $8\%$ at most for the cases considered here.

Additionally, we show in the inset of Fig.~\ref{fig:uni_rel} a comparison
of the linear and nonlinear fittings for the threshold mass in the whole
range of possible compactnesses, \ie $\mathcal{C}_{_{\rm
    TOV}}\in[0,1/2]$. The light-green area refers to neutron stars, with
the lower limit $\mathcal{C}_{_{\rm TOV}}\gtrsim0.2$ deduced from a large
sample of EOSs \citep{Most2018} and the upper limit $\mathcal{C}_{_{\rm
    TOV}}\lesssim0.35$ set by causality; instead, the upper limit of the
dark-green area $\mathcal{C}_{_{\rm TOV}}<4/9$ is set by the Buchdahl
limit for compact stars~\citep{Rezzolla_book:2013}.

\begin{figure*}
	\centering
	\includegraphics[width=\columnwidth]{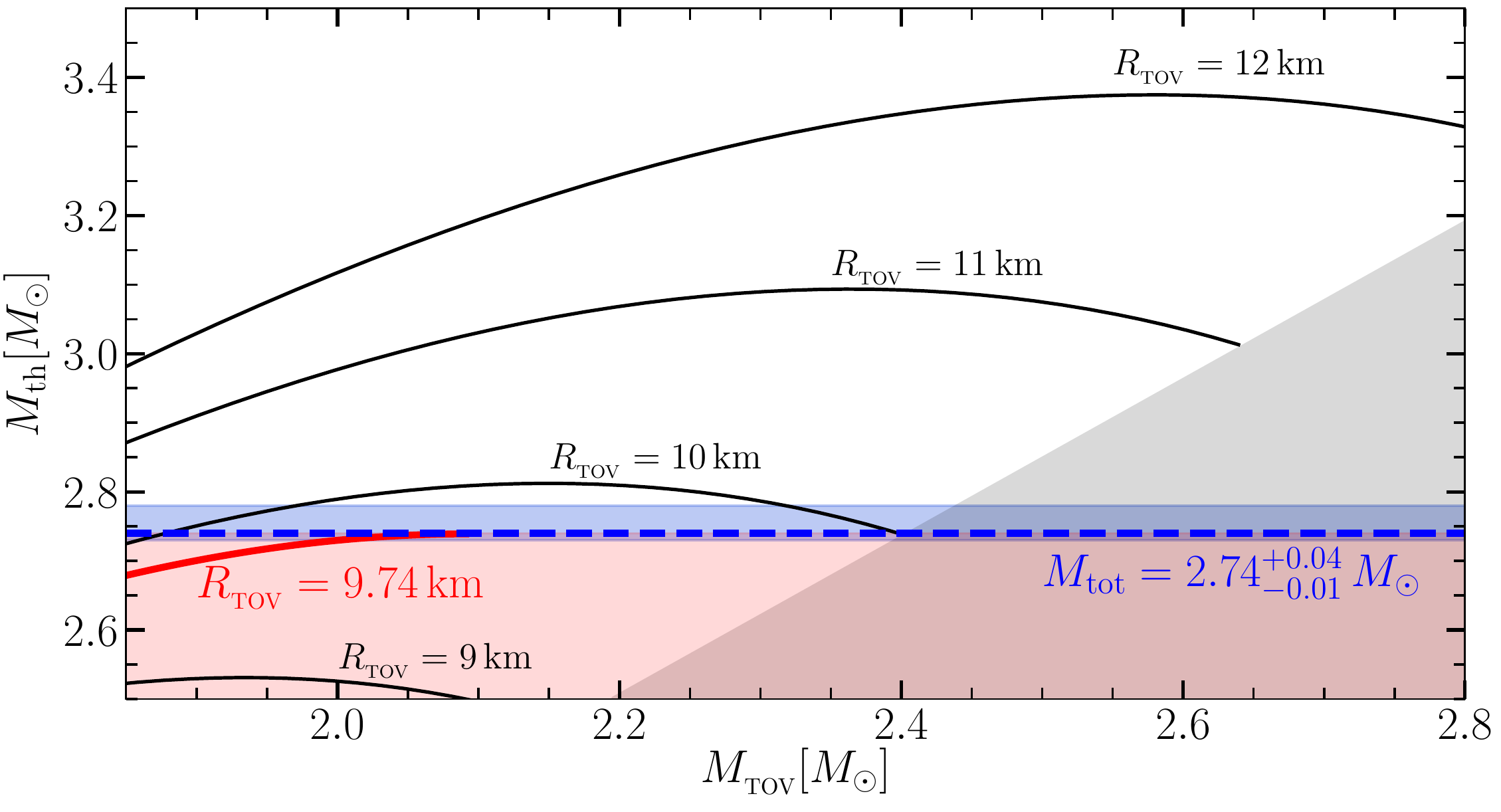}
        \hskip 0.5cm
	\includegraphics[width=\columnwidth]{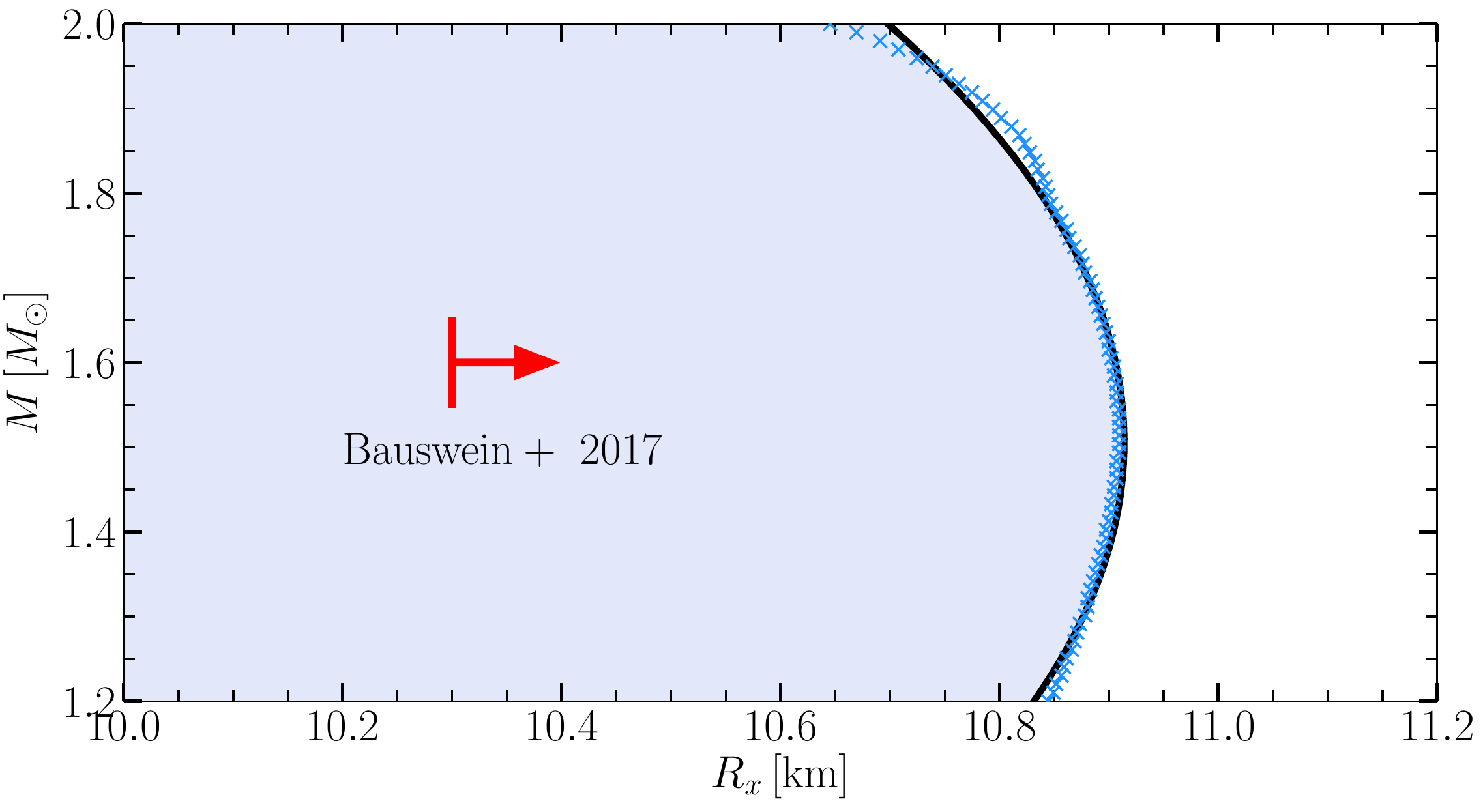}
	\caption{\textit{Left panel:} the lower bound on $R_{_{\rm TOV}}$
          (red) using the universal relation (\ref{eq:fit}). The
          horizontal blue-dashed line marks the mass of GW170817 and its
          uncertainty. The red-shaded area shows the values excluded by
          the detection. The grey-shaded area represents values excluded
          by the causality constraint. \textit{Right panel:} universal
          relation (black) for the lower limit of $R_{x}$ for a given
          mass $M$ (blue crossed); the red arrow represents the
          constraint from~\citet{Bauswein2017b} for a $1.6 M_{\odot}$
          star.}
	\label{fig:radius-constraint}
\end{figure*}

Besides providing an improvement over the linear approximation of
\citet{Bauswein2013}, the nonlinear expression~\eqref{eq:fit} can now be
used to provide more stringent (\ie larger) lower limits on the radii of
possible stellar models in the light of the recent detection of
GW170817~\citep{Abbott2017_etal}. In particular, following
\citet{Bauswein2017b}, we use \eqref{eq:fit} to plot the threshold mass
for different potential choices of $R_{_{\rm TOV}}$; this is shown in the
left panel of Fig.~\ref{fig:radius-constraint} with black solid lines and
for $R_{_{\rm TOV}}=10, 11, 12\,{\rm km}$. Also reported in
Fig.~\ref{fig:radius-constraint} with a grey-shaded area is the limit set
by causality and that requires $M_{_{\rm TOV}}/R_{_{\rm
    TOV}}\lesssim0.354$~\citep{Koranda1997}. As noted
by~\citet{Bauswein2017b}, given the merger of a neutron-star binary with
total mass $M_{\rm tot}$, it is possible to set a lower limit on $M_{\rm
  th}$. This is shown in the left panel Fig.~\ref{fig:radius-constraint},
where we report with a horizontal blue-dashed line the total
gravitational mass estimated for GW170817, $M_{\rm
  tot}=2.74^{+0.04}_{-0.01}\,M_{\odot}$~\citep{Abbott2017_etal}.

The corresponding uncertainty band (blue-shaded area) gives a lower
constraint on $M_{\rm th}$, since GW170817 did not lead to a prompt
collapse. The blue band thus constraints the red-shaded area from below,
yielding a lower limit for the radius of the maximum-mass star, $R_{_{\rm
    TOV}} \ge 9.74^{+0.14}_{-0.04}\,{\rm km}$ (red solid line); this is
to be contrasted with the value deduced by~\citet{Bauswein2017b}, \ie
$R_{_{\rm TOV}} \ge 9.26^{+0.17}_{-0.03}\,{\rm km}$, on the basis of
their linear approximation. Interestingly, to obtain a more stringent
constraint similar to the one derived here,~\citet{Bauswein2017b}
required a hypothetical detection of a binary with $M_{\rm tot} \simeq
2.9\,M_{\odot}$.

The logical approach and the mathematical procedure followed so far to
derive the nonlinear fit \eqref{eq:fit} for $\mathcal{C}_{_{\rm TOV}}$
can be repeated for the compactness of a fixed mass $M_x$, \ie
$\mathcal{C}_{x}:=M_x/R_x$, thus allowing us to set constraints not only
on $R_{_{\rm TOV}}$, but on any radius $R_x$ within a reasonable range
(the fit becomes increasingly bad for large masses since the EOSs do not
have solutions well above two solar masses). This is shown in the right
panel of Fig.~\ref{fig:radius-constraint}, where the values of $M_x$ and
$R_x$ are indicated with blue crosses, while in black is the quadratic
fit
\begin{equation}
\label{eq:fit_generic}
R_x = -0.88\,M^{2} + 2.66\,M + 8.91\,.
\end{equation}

The importance of expression \eqref{eq:fit_generic} is that it offers a
handy expression for the lower limit of stellar models as deduced from
GW170817. A similar procedure has been followed by \citet{Bauswein2017b},
but only for a fixed mass of $1.6\,M_{\odot}$, from which it was deduced
that $R_{1.6}\ge 10.30\,{\rm km}$; this result should be contrasted with
the value derived from \eqref{eq:fit_generic}, which is instead
$R_{1.6}\ge10.90\,{\rm km}$.  Similarly, for a reference star of
$1.4\,M_{\odot}$ we obtain $R_{1.4}\ge10.92\,{\rm km}$, which is
worryingly close to the estimate by~\citet{Bauswein2017b} for
$1.6\,M_{\odot}$\footnote{\citet{Bauswein2017b} provide an estimate only
  for $R_{1.6}$.}. On the other hand, our $R_{1.4}$ estimate is in good
agreement with that of \citet{Most2018}, who have explored a large number
of possible EOSs and built a set of one billion stellar models from which
they deduced that $12.00<R_{1.4}/{\rm km}<13.45$.

\section{Conclusions}
\label{sec:conclusions}

The detection of the merger of a binary system of neutron stars has
concretely initiated the process of extracting information on the most
extreme state of matter from gravitational-wave signals. An important
phase of this process lies in understanding the post-merger behaviour of
the binary and the stability of the remnant. Indeed, knowing whether the
system promptly forms a black hole is critical to understanding and
interpreting the electromagnetic signals that may be observed.

Using a fully general-relativistic approach and a novel method for the
determination of the threshold mass, we have carried out simulations
making use of all of the realistic EOSs available to describe this
process. In this way, we have found a \emph{nonlinear} universal relation
for the threshold mass as a function of the maximum compactness and which
is potentially valid for all compactnesses. At least for the
temperature-dependent EOSs considered here, this universal relation
improves the \emph{linear} relation found recently with methods that are
less accurate, but that also yield quantitatively similar
results. Furthermore, exploiting the detection of GW170817, we have used
the universal relation to set lower limits on the stellar radii for any
any mass.

These results can be improved in at least two ways. First, as new hot
EOSs becomes available for numerical simulations it will be possible to
extend the analysis carried here, reducing its uncertainty. Second, as
new detections from binary neutron-star mergers will be revealed, the
masses of these systems and their electromagnetic counterparts will be
used to set ever more precise lower bounds on the radii of neutron stars.

\section*{Acknowledgements}

Support comes from: the ERC synergy grant ``BlackHoleCam: Imaging the
Event Horizon of Black Holes'' (Grant No. 610058), ``PHAROS'', COST
Action CA16214; the LOEWE-Program in HIC for FAIR; the European Union's
Horizon 2020 Research and Innovation Programme (Grant 671698) (call
FETHPC-1-2014, project ExaHyPE). The simulations were performed on the
clusters SuperMUC (LRZ, Garching), LOEWE (CSC, Frankfurt), and HazelHen
(HLRS, Stuttgart). The EOSs employed can be found on
\href{https://stellarcollapse.org}{stellarcollapse.org}.



\begin{thebibliography}{}
\providecommand\natexlab[1]{#1}
\providecommand\JournalTitle[1]{#1}

\bibitem[{Abbott {et~al.}(2017)}]{Abbott2017_etal}
Abbott, B.~P., {et~al.} 2017,
  \href{http://dx.doi.org/10.1103/PhysRevLett.119.161101}{\JournalTitle{Phys.
  Rev. Lett.}, 119, 161101}

\bibitem[{Alcubierre(2008)}]{Alcubierre:2008}
Alcubierre, M. 2008, Introduction to $3+1$ {N}umerical {R}elativity (Oxford,
  UK: Oxford University Press)

\bibitem[{{Alford} {et~al.}(2018){Alford}, {Bovard}, {Hanauske}, {Rezzolla}, \&
  {Schwenzer}}]{Alford2017}
{Alford}, M.~G., {Bovard}, L., {Hanauske}, M., {Rezzolla}, L., \& {Schwenzer},
  K. 2018,
  \href{http://dx.doi.org/10.1103/PhysRevLett.120.041101}{\JournalTitle{Phys.
  Rev. Lett.}, 120, 041101}

\bibitem[{{Annala} {et~al.}(2018){Annala}, {Gorda}, {Kurkela}, \&
  {Vuorinen}}]{Annala2017}
{Annala}, E., {Gorda}, T., {Kurkela}, A., \& {Vuorinen}, A. 2018,
  \href{http://dx.doi.org/10.1103/PhysRevLett.120.172703}{\JournalTitle{Phys.
  Rev. Lett.}, 120, 172703}

\bibitem[{{Baiotti} {et~al.}(2008){Baiotti}, {Giacomazzo}, \&
  {Rezzolla}}]{Baiotti08}
{Baiotti}, L., {Giacomazzo}, B., \& {Rezzolla}, L. 2008,
  \href{http://dx.doi.org/10.1103/PhysRevD.78.084033}{\JournalTitle{Phys. Rev.
  D}, 78, 084033}

\bibitem[{Baiotti \& Rezzolla(2017)}]{Baiotti2016}
Baiotti, L., \& Rezzolla, L. 2017,
  \href{http://dx.doi.org/10.1088/1361-6633/aa67bb}{\JournalTitle{Rept. Prog.
  Phys.}, 80, 096901}

\bibitem[{{Banik} {et~al.}(2014){Banik}, {Hempel}, \&
  {Bandyopadhyay}}]{Banik2014}
{Banik}, S., {Hempel}, M., \& {Bandyopadhyay}, D. 2014,
  \href{http://dx.doi.org/10.1088/0067-0049/214/2/22}{\JournalTitle{Astrohys.
  J. Suppl.}, 214, 22}

\bibitem[{{Bauswein} {et~al.}(2013){Bauswein}, {Baumgarte}, \&
  {Janka}}]{Bauswein2013}
{Bauswein}, A., {Baumgarte}, T.~W., \& {Janka}, H.-T. 2013,
  \href{http://dx.doi.org/10.1103/PhysRevLett.111.131101}{\JournalTitle{Phys.
  Rev. Lett.}, 111, 131101}

\bibitem[{{Bauswein} {et~al.}(2017){Bauswein}, {Just}, {Janka}, \&
  {Stergioulas}}]{Bauswein2017b}
{Bauswein}, A., {Just}, O., {Janka}, H.-T., \& {Stergioulas}, N. 2017,
  \href{http://dx.doi.org/10.3847/2041-8213/aa9994}{\JournalTitle{Astrophys. J.
  Lett.}, 850, L34}

\bibitem[{{Bovard} {et~al.}(2017){Bovard}, {Martin}, {Guercilena}, {Arcones},
  {Rezzolla}, \& {Korobkin}}]{Bovard2017}
{Bovard}, L., {Martin}, D., {Guercilena}, F., {et~al.} 2017,
  \JournalTitle{Phys. Rev. D}, 96, 124005

\bibitem[{{Breu} \& {Rezzolla}(2016)}]{Breu2016}
{Breu}, C., \& {Rezzolla}, L. 2016,
  \href{http://dx.doi.org/10.1093/mnras/stw575}{\JournalTitle{Mon. Not. R.
  Astron. Soc.}, 459, 646}

\bibitem[{Brown {et~al.}(2009)Brown, Diener, Sarbach, Schnetter, \&
  Tiglio}]{Brown:2008sb}
Brown, D., Diener, P., Sarbach, O., Schnetter, E., \& Tiglio, M. 2009,
  \href{http://dx.doi.org/10.1103/PhysRevD.79.044023}{\JournalTitle{Phys. Rev.
  D}, 79, 044023}

\bibitem[{{Burgio} {et~al.}(2018){Burgio}, {Drago}, {Pagliara}, {Schulze}, \&
  {Wei}}]{Burgio2018}
{Burgio}, G.~F., {Drago}, A., {Pagliara}, G., {Schulze}, H.-J., \& {Wei}, J.-B.
  2018,
  \href{http://dx.doi.org/10.3847/1538-4357/aac6ee}{\JournalTitle{Astrophys.
  J.}, 860, 139}

\bibitem[{{Dietrich} \& {Ujevic}(2017)}]{Dietrich2016}
{Dietrich}, T., \& {Ujevic}, M. 2017,
  \href{http://dx.doi.org/10.1088/1361-6382/aa6bb0}{\JournalTitle{Classical and
  Quantum Gravity}, 34, 105014}

\bibitem[{{Duez} {et~al.}(2004){Duez}, {Liu}, {Shapiro}, \&
  {Stephens}}]{Duez2004b}
{Duez}, M.~D., {Liu}, Y.~T., {Shapiro}, S.~L., \& {Stephens}, B.~C. 2004,
  \href{http://dx.doi.org/10.1103/PhysRevD.69.104030}{\JournalTitle{Phys. Rev.
  D}, 69, 104030}

\bibitem[{{Eichler} {et~al.}(1989){Eichler}, {Livio}, {Piran}, \&
  {Schramm}}]{Eichler89}
{Eichler}, D., {Livio}, M., {Piran}, T., \& {Schramm}, D.~N. 1989,
  \href{http://dx.doi.org/10.1038/340126a0}{\JournalTitle{Nature}, 340, 126}

\bibitem[{{Gourgoulhon} {et~al.}(2001){Gourgoulhon}, {Grandcl{\'e}ment},
  {Taniguchi}, {Marck}, \&
  {Bonazzola}}]{Gourgoulhon-etal-2000:2ns-initial-data}
{Gourgoulhon}, E., {Grandcl{\'e}ment}, P., {Taniguchi}, K., {Marck}, J.-A., \&
  {Bonazzola}, S. 2001,
  \href{http://dx.doi.org/10.1103/PhysRevD.63.064029}{\JournalTitle{Phys. Rev.
  D}, 63, 064029}

\bibitem[{{Hanauske} {et~al.}(2017){Hanauske}, {Takami}, {Bovard}, {Rezzolla},
  {Font}, {Galeazzi}, \& {St{\"o}cker}}]{Hanauske2016}
{Hanauske}, M., {Takami}, K., {Bovard}, L., {et~al.} 2017,
  \href{http://dx.doi.org/10.1103/PhysRevD.96.043004}{\JournalTitle{Phys. Rev.
  D}, 96, 043004}

\bibitem[{{Hempel} {et~al.}(2012){Hempel}, {Fischer}, {Schaffner-Bielich}, \&
  {Liebend{\"o}rfer}}]{Hempel2012}
{Hempel}, M., {Fischer}, T., {Schaffner-Bielich}, J., \& {Liebend{\"o}rfer}, M.
  2012,
  \href{http://dx.doi.org/10.1088/0004-637X/748/1/70}{\JournalTitle{Astrophys.
  J.}, 748, 70}

\bibitem[{Janka {et~al.}(1993)Janka, Zwerger, \& M{\"o}nchmeyer}]{Janka93}
Janka, H.-T., Zwerger, T., \& M{\"o}nchmeyer, R. 1993, \JournalTitle{Astron.
  Astrophys.}, 268, 360

\bibitem[{{Kastaun} {et~al.}(2016){Kastaun}, {Ciolfi}, \&
  {Giacomazzo}}]{Kastaun2016}
{Kastaun}, W., {Ciolfi}, R., \& {Giacomazzo}, B. 2016,
  \href{http://dx.doi.org/10.1103/PhysRevD.94.044060}{\JournalTitle{Phys. Rev.
  D}, 94, 044060}

\bibitem[{{Kawamura} {et~al.}(2016){Kawamura}, {Giacomazzo}, {Kastaun},
  {Ciolfi}, {Endrizzi}, {Baiotti}, \& {Perna}}]{Kawamura2016}
{Kawamura}, T., {Giacomazzo}, B., {Kastaun}, W., {et~al.} 2016,
  \href{http://dx.doi.org/10.1103/PhysRevD.94.064012}{\JournalTitle{Phys. Rev.
  D}, 94, 064012}

\bibitem[{{Kiuchi} {et~al.}(2015){Kiuchi}, {Cerd{\'a}-Dur{\'a}n}, {Kyutoku},
  {Sekiguchi}, \& {Shibata}}]{Kiuchi2015a}
{Kiuchi}, K., {Cerd{\'a}-Dur{\'a}n}, P., {Kyutoku}, K., {Sekiguchi}, Y., \&
  {Shibata}, M. 2015,
  \href{http://dx.doi.org/10.1103/PhysRevD.92.124034}{\JournalTitle{Phys. Rev.
  D}, 92, 124034}

\bibitem[{{Koranda} {et~al.}(1997){Koranda}, {Stergioulas}, \&
  {Friedman}}]{Koranda1997}
{Koranda}, S., {Stergioulas}, N., \& {Friedman}, J.~L. 1997,
  \href{http://dx.doi.org/10.1086/304714}{\JournalTitle{Astrophys. J.}, 488,
  799}

\bibitem[{{Kyutoku} {et~al.}(2014){Kyutoku}, {Ioka}, \&
  {Shibata}}]{Kyutoku2012}
{Kyutoku}, K., {Ioka}, K., \& {Shibata}, M. 2014,
  \href{http://dx.doi.org/10.1093/mnrasl/slt128}{\JournalTitle{Mon. Not. R.
  Astron.Soc.}, 437, L6}

\bibitem[{{Lehner} {et~al.}(2016){Lehner}, {Liebling}, {Palenzuela},
  {Caballero}, {O'Connor}, {Anderson}, \& {Neilsen}}]{Lehner2016}
{Lehner}, L., {Liebling}, S.~L., {Palenzuela}, C., {et~al.} 2016,
  \href{http://dx.doi.org/10.1088/0264-9381/33/18/184002}{\JournalTitle{Classical
  and Quantum Gravity}, 33, 184002}

\bibitem[{{L{\"o}ffler} {et~al.}(2012){L{\"o}ffler}, {Faber}, {Bentivegna},
  {Bode}, {Diener}, {Haas}, {Hinder}, {Mundim}, {Ott}, {Schnetter}, {Allen},
  {Campanelli}, \& {Laguna}}]{loeffler_2011_et}
{L{\"o}ffler}, F., {Faber}, J., {Bentivegna}, E., {et~al.} 2012,
  \href{http://dx.doi.org/10.1088/0264-9381/29/11/115001}{\JournalTitle{Class.
  Quantum Grav.}, 29, 115001}

\bibitem[{{Margalit} \& {Metzger}(2017)}]{Margalit2017}
{Margalit}, B., \& {Metzger}, B.~D. 2017,
  \href{http://dx.doi.org/10.3847/2041-8213/aa991c}{\JournalTitle{Astrophys. J.
  Lett.}, 850, L19}

\bibitem[{{Montana} {et~al.}(2018){Montana}, {Tolos}, {Hanauske}, \&
  {Rezzolla}}]{Montana2018}
{Montana}, G., {Tolos}, L., {Hanauske}, M., \& {Rezzolla}, L. 2018,
  \JournalTitle{arXiv:1811.10929}, arXiv:1811.10929

\bibitem[{{Most} {et~al.}(2018){Most}, {Weih}, {Rezzolla}, \&
  {Schaffner-Bielich}}]{Most2018}
{Most}, E.~R., {Weih}, L.~R., {Rezzolla}, L., \& {Schaffner-Bielich}, J. 2018,
  \href{http://dx.doi.org/10.1103/PhysRevLett.120.261103}{\JournalTitle{Phys.
  Rev. Lett.}, 120, 261103}

\bibitem[{{Paschalidis} {et~al.}(2017){Paschalidis}, {Yagi},
  {Alvarez-Castillo}, {Blaschke}, \& {Sedrakian}}]{Paschalidis2017}
{Paschalidis}, V., {Yagi}, K., {Alvarez-Castillo}, D., {Blaschke}, D.~B., \&
  {Sedrakian}, A. 2017, \JournalTitle{arXiv:1712.00451},
  \href{http://arxiv.org/abs/1712.00451}{{\sffamily arXiv:1712.00451
  [astro-ph.HE]}}

\bibitem[{{Radice}(2017)}]{Radice2017}
{Radice}, D. 2017,
  \href{http://dx.doi.org/10.3847/2041-8213/aa6483}{\JournalTitle{Astrophys. J.
  Lett.}, 838, L2}

\bibitem[{{Radice} {et~al.}(2016){Radice}, {Galeazzi}, {Lippuner}, {Roberts},
  {Ott}, \& {Rezzolla}}]{Radice2016}
{Radice}, D., {Galeazzi}, F., {Lippuner}, J., {et~al.} 2016,
  \href{http://dx.doi.org/10.1093/mnras/stw1227}{\JournalTitle{Mon. Not. R.
  Astron. Soc.}, 460, 3255}

\bibitem[{{Radice} {et~al.}(2018){Radice}, {Perego}, {Zappa}, \&
  {Bernuzzi}}]{Radice2017b}
{Radice}, D., {Perego}, A., {Zappa}, F., \& {Bernuzzi}, S. 2018,
  \href{http://dx.doi.org/10.3847/2041-8213/aaa402}{\JournalTitle{Astrophys. J.
  Lett.}, 852, L29}

\bibitem[{{Radice} \& {Rezzolla}(2012)}]{Radice2012a}
{Radice}, D., \& {Rezzolla}, L. 2012,
  \href{http://dx.doi.org/10.1051/0004-6361/201219735}{\JournalTitle{Astron.
  Astrophys.}, 547, A26}

\bibitem[{{Radice} {et~al.}(2014){Radice}, {Rezzolla}, \&
  {Galeazzi}}]{Radice2013c}
{Radice}, D., {Rezzolla}, L., \& {Galeazzi}, F. 2014,
  \href{http://dx.doi.org/10.1088/0264-9381/31/7/075012}{\JournalTitle{Class.
  Quantum Grav.}, 31, 075012}

\bibitem[{{Rezzolla} {et~al.}(2011){Rezzolla}, {Giacomazzo}, {Baiotti},
  {Granot}, {Kouveliotou}, \& {Aloy}}]{Rezzolla:2011}
{Rezzolla}, L., {Giacomazzo}, B., {Baiotti}, L., {et~al.} 2011,
  \href{http://dx.doi.org/10.1088/2041-8205/732/1/L6}{\JournalTitle{Astrophys.
  J. Letters}, 732, L6}

\bibitem[{{Rezzolla} {et~al.}(2018){Rezzolla}, {Most}, \&
  {Weih}}]{Rezzolla2017}
{Rezzolla}, L., {Most}, E.~R., \& {Weih}, L.~R. 2018,
  \href{http://dx.doi.org/10.3847/2041-8213/aaa401}{\JournalTitle{Astrophys. J.
  Lett.}, 852, L25}

\bibitem[{{Rezzolla} \& {Zanotti}(2013)}]{Rezzolla_book:2013}
{Rezzolla}, L., \& {Zanotti}, O. 2013, Relativistic Hydrodynamics (Oxford, UK:
  Oxford University Press)

\bibitem[{{Rosswog} {et~al.}(1999){Rosswog}, {Liebend{\"o}rfer}, {Thielemann},
  {Davies}, {Benz}, \& {Piran}}]{Rosswog1999}
{Rosswog}, S., {Liebend{\"o}rfer}, M., {Thielemann}, F.-K., {et~al.} 1999,
  \JournalTitle{Astron. Astrophys.}, 341, 499

\bibitem[{{Ruiz} {et~al.}(2018){Ruiz}, {Shapiro}, \& {Tsokaros}}]{Ruiz2017}
{Ruiz}, M., {Shapiro}, S.~L., \& {Tsokaros}, A. 2018,
  \href{http://dx.doi.org/10.1103/PhysRevD.97.021501}{\JournalTitle{Phys. Rev.
  D}, 97, 021501}

\bibitem[{{Schnetter} {et~al.}(2004){Schnetter}, {Hawley}, \&
  {Hawke}}]{Schnetter-etal-03b}
{Schnetter}, E., {Hawley}, S.~H., \& {Hawke}, I. 2004,
  \href{http://dx.doi.org/10.1088/0264-9381/21/6/014}{\JournalTitle{Class.
  Quantum Grav.}, 21, 1465}

\bibitem[{{Shibata} {et~al.}(2017){Shibata}, {Fujibayashi}, {Hotokezaka},
  {Kiuchi}, {Kyutoku}, {Sekiguchi}, \& {Tanaka}}]{Shibata2017c}
{Shibata}, M., {Fujibayashi}, S., {Hotokezaka}, K., {et~al.} 2017,
  \href{http://dx.doi.org/10.1103/PhysRevD.96.123012}{\JournalTitle{Phys. Rev.
  D}, 96, 123012}

\bibitem[{{Shibata} \& {Kiuchi}(2017)}]{Shibata:2017b}
{Shibata}, M., \& {Kiuchi}, K. 2017,
  \href{http://dx.doi.org/10.1103/PhysRevD.95.123003}{\JournalTitle{Phys. Rev.
  D}, 95, 123003}

\bibitem[{{Siegel} {et~al.}(2013){Siegel}, {Ciolfi}, {Harte}, \&
  {Rezzolla}}]{Siegel2013}
{Siegel}, D.~M., {Ciolfi}, R., {Harte}, A.~I., \& {Rezzolla}, L. 2013,
  \href{http://dx.doi.org/10.1103/PhysRevD.87.121302}{\JournalTitle{Phys. Rev.
  D R}, 87, 121302}

\bibitem[{{Steiner} {et~al.}(2013){Steiner}, {Hempel}, \&
  {Fischer}}]{Steiner2013}
{Steiner}, A.~W., {Hempel}, M., \& {Fischer}, T. 2013,
  \href{http://dx.doi.org/10.1088/0004-637X/774/1/17}{\JournalTitle{Astrophys.
  J.}, 774, 17}

\bibitem[{{Tews} {et~al.}(2017){Tews}, {Lattimer}, {Ohnishi}, \&
  {Kolomeitsev}}]{Kolomeitsev2016}
{Tews}, I., {Lattimer}, J.~M., {Ohnishi}, A., \& {Kolomeitsev}, E.~E. 2017,
  \href{http://dx.doi.org/10.3847/1538-4357/aa8db9}{\JournalTitle{Astrophys.
  J.}, 848, 105}

\bibitem[{Togashi {et~al.}(2016)Togashi, Hiyama, Yamamoto, \&
  Takano}]{Togashi2016}
Togashi, H., Hiyama, E., Yamamoto, Y., \& Takano, M. 2016,
  \href{http://dx.doi.org/10.1103/PhysRevC.93.035808}{\JournalTitle{Phys.
  Rev.}, C93, 035808}

\bibitem[{{Typel} {et~al.}(2010){Typel}, {R{\"o}pke}, {Kl{\"a}hn}, {Blaschke},
  \& {Wolter}}]{Typel2010}
{Typel}, S., {R{\"o}pke}, G., {Kl{\"a}hn}, T., {Blaschke}, D., \& {Wolter},
  H.~H. 2010,
  \href{http://dx.doi.org/10.1103/PhysRevC.81.015803}{\JournalTitle{Phys. Rev.
  C}, 81, 015803}

\bibitem[{{Weih} {et~al.}(2018){Weih}, {Most}, \& {Rezzolla}}]{Weih2017}
{Weih}, L.~R., {Most}, E.~R., \& {Rezzolla}, L. 2018,
  \href{http://dx.doi.org/10.1093/mnrasl/slx178}{\JournalTitle{Mon. Not. R.
  Astron. Soc.}, 473, L126}

\end{thebibliography}
\end{document}